\documentclass{iau} 
\usepackage{graphicx}

\title[Toward a full Molecular Cloud catalog of the Galaxy] 
{The road toward a full, high resolution \\Molecular Cloud catalog of the Galaxy}

\author[Dario Colombo et al.]   
{Dario Colombo$^1$,
Erik Rosolowsky$^1$,
Adam Ginsburg$^2$,\\
Ana Duarte-Cabral$^3$,
\and Annie Hughes$^{4,5,6}$}

\affiliation{$^1$ Department of Physics, University of Alberta, \\ 4-181 CCIS, Edmonton, AB T6G 2E1, Canada, email: {\tt dcolombo@ualberta.ca} \\[\affilskip]
$^2$ European Southern Observatory, Karl-Schwarzschild-Strasse 2, \\ 85748, Garching bei Munchen, Germany \\[\affilskip]
$^3$ School of Physics and Astronomy, University of Exeter, Stocker Road, Exeter EX4 4QL, UK \\[\affilskip]
$^4$ CNRS, IRAP, 9 Av. colonel Roche, BP 44346, F-31028 Toulouse cedex 4, France \\[\affilskip]
$^5$ Universit\'{e} de Toulouse, UPS-OMP, IRAP, F-31028 Toulouse cedex 4, France \\[\affilskip]
$^6$ Max Planck Institute for Astronomy, K\"onigstuhl 17, 69117 Heidelberg, Germany}

\pubyear{2015}
\volume{xxx}  
\jname{From Interstellar Clouds to Star-forming Galaxies: \\Universal Processes?}
\editors{A.C. Editor, B.D. Editor \& C.E. Editor, eds.}
\begin{document}

\maketitle

\begin{abstract}
The statistical description of Giant Molecular Cloud (GMC) properties relies heavily on the performance of automatic identification algorithms, which are often seriously affected by the survey design. The algorithm we designed, SCIMES (Spectral Clustering for Interstellar Molecular Emission Segmentation), is able to overcome some of these limitations by considering the cloud segmentation problem in the broad framework of the graph theory. The application of the code on the CO(3-2) High Resolution Survey (COHRS) data allowed for a robust decomposition of more than 12,000 objects in the Galactic Plane. Together with the wealth of Galactic Plane surveys of the recent years, this approach will help to open the door to a future, systematic cataloging of all discrete molecular features of our own Galaxy.
\end{abstract}

{\bf Introduction.} Since the first large surveys of early 1980s, the intrinsic clumpiness of the molecular interstellar medium (ISM, \cite[Leroy et al. 2013]{leroy13}) drove the cataloging of those clumps, the larger of which (typically 40\,pc and 10$^5$\,M$_{\odot}$) have been called GMCs (\cite[Solomon et al. 1979]{solomon87}). The statistical description of GMC properties has provided important insights into the physics of the star formation. GMCs appear roughly gravitationally bound and present a hierarchical structure (e.g., \cite[Solomon et al. 1987]{solomon87}). These observations suggest that clouds are sustained against gravitational collapse by turbulence, which might also explain the low star formation efficiencies observed in the Galaxy (e.g., \cite[Mckee \& Ostriker 2007]{mckee07}).Those conclusions have been mostly obtained by applying automatic algorithms to decouple the discrete ISM features from the more diffuse intra-cloud medium. Unfortunately those algorithms struggle to recognize GMCs as single entities once they span several resolution elements (as for the most recent high resolution Galactic Plane surveys). The algorithm we designed (SCIMES, Colombo et al., MNRAS in press) by applying the spectral clustering paradigm to dendrograms of molecular gas emission (see \cite[Rosolowsky et al. 2008]{rosolowsky08}) appears to be able to overcome such difficulties. The spectral clustering technique selects the relevant objects within the dendrogram, based on the clustering properties of ISM chosen and the scale of the observation. SCIMES might also be used to find other molecular gas structure (as clumps, filaments, and ``Giant Molecular Associations''), given enough dynamical range to define their hierarchy and the correct set of clustering criteria. In this aspect, every ISM structure might be seen as a sub-class of a more extended class of objects: the ``Molecular Gas Clusters''.\\  



\begin{figure}[ht]
\begin{center}
 \includegraphics[width=\textwidth]{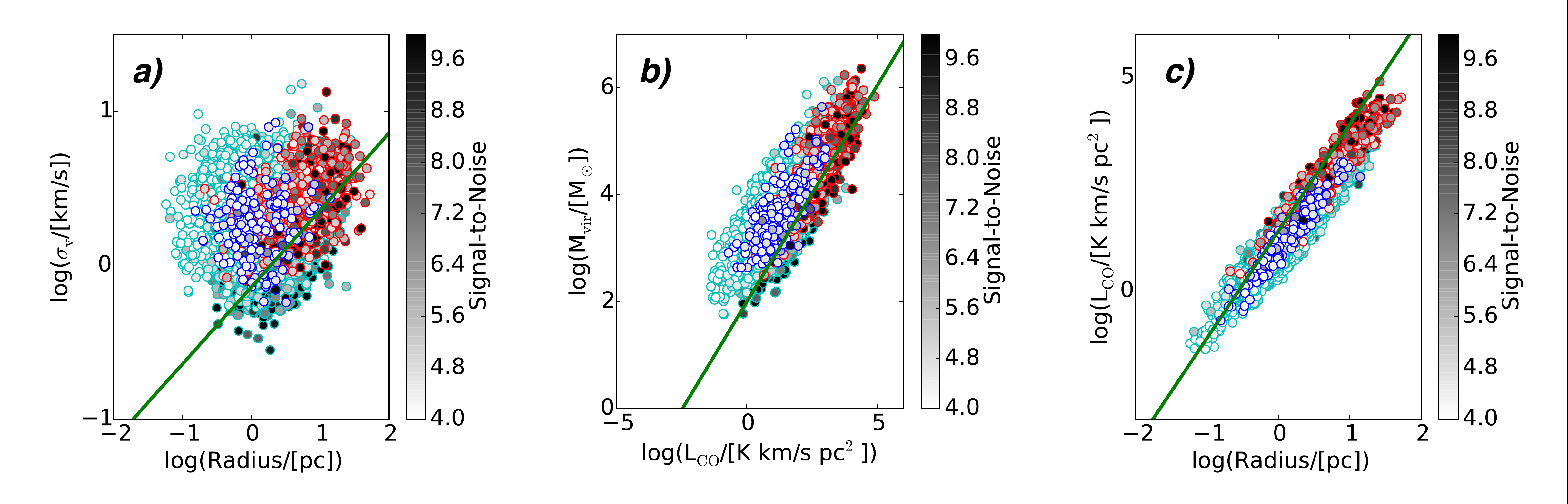} 
 \caption{COHRS cloud scaling relations. Red circles indicate clouds with a well defined distance, blue circles mark objects where the distance point falls within one of the parental structures, while cyan circles represent isolated objects whose distance is given by the closest distance point in the PPV space. Green lines indicate the
fits by \cite[Solomon et al. 1987]{solomon87}.}
   \label{cohrs}
\end{center}
\end{figure}

{\bf The COHRS cloud catalog.} We applied SCIMES on the COHRS data (\cite[Dempsey et al. 2013]{dempsey13}). COHRS mapped the $^{12}$CO(3-2) emission from the inner Galactic Plane ($b\leq|0.5^\circ|$, $10.25^\circ\leq l \leq 55.25^\circ$, $-30 \leq v_{\mathrm{LSR}} \leq 155$ km/s) with a spatial and spectral resolution of 16.6" and 1 km/s, respectively. We cataloged 12,641 objects in total. The dendrogram together with the clustering approach allows to automatically identify a variety of gas morphologies including coherent filamentary structures and holes within the molecular ISM. We generate physical properties of the clouds using the high quality distances obtained by the Bolocam Galactic Plane Survey (\cite[Ellsworth-Bowers et al. 2015]{ellsworth15}) and we plot those properties against each other to probe the validity of the ``Larson's laws'' within this sample. We do not observe a clear size-line width relation (Fig~\ref{cohrs}a), but the scatter is introduced mostly by uncertain distance clouds. Nevertheless, clouds appear to have, on average, similar surface brightness as predicted (Fig~\ref{cohrs}b). Given the low scatter in the CO luminosity-virial mass relation (Fig~\ref{cohrs}c), $X_{\mathrm{CO}}$ does not vary much from the adopted Galactic value: $X_{\mathrm{CO(3-2)}} = 5\times10^{20}$\,cm$^{-2}$\,K$^{-1}$\,km$^{-1}$ (in a picture consistent with self-
gravitating clouds). In the future we will study how the properties, the morphology, and the turbulence of these clouds influence their star formation capability. This analysis (Colombo et al. in preparation) will give the guidelines to create the full, high resolution molecular cloud catalog of entire Galactic Plane.

\end{document}